\documentclass[conference]{IEEEtran}

\usepackage{graphicx}
\usepackage{array}
\usepackage{tabularx}
\usepackage[flushleft]{threeparttable}
\usepackage{cite}

\usepackage[absolute]{textpos}

\begin{document}

\begin{textblock}{13}(1.5,0.2)
{\footnotesize \noindent U. De Silva, T. Koike-Akino, R. Ma,  A. Yamashita, and H. Nakamizo, ``A Modular 1D-CNN Architecture for Real-time Digital Pre-distortion," \\ \textit{RWW2022: IEEE Radio \& Wireless Week}, Las Vegas, NV, USA, January 2022, pp. 1-3.}
\end{textblock}

%
\title{A Modular 1D-CNN Architecture for \\Real-time Digital Pre-distortion }

\author{\IEEEauthorblockN{Udara De Silva\IEEEauthorrefmark{1}, Toshiaki Koike-Akino\IEEEauthorrefmark{1}, Rui Ma\IEEEauthorrefmark{1}, Ao Yamashita\IEEEauthorrefmark{2}, and Hideyuki Nakamizo\IEEEauthorrefmark{2}}

\IEEEauthorblockA{\IEEEauthorrefmark{1}Mitsubishi Electric Research Labs, Cambridge, MA, USA, rma@merl.com}

\IEEEauthorblockA{
\IEEEauthorrefmark{2}Mitsubishi Electric Corporation, Information Tech. R\&D Center, Kanagawa, Japan
}}


%


\maketitle

\begin{abstract}
This study reports a novel hardware-friendly modular architecture for implementing one dimensional convolutional neural network (1D-CNN) digital predistortion (DPD) technique to linearize RF power amplifier (PA) real-time. 
The modular nature of our design enables DPD system adaptation for variable resource and timing constraints. 
Our work also presents a co-simulation architecture to verify the DPD performance with an actual power amplifier hardware-in-the-loop.
The experimental results with 100 MHz signals show that the proposed 1D-CNN obtains superior performance compared with other neural network architectures for real-time DPD application. 
\end{abstract}


%
\IEEEpeerreviewmaketitle

\section{Introduction}

The fifth-generation (5G) new radio (NR) is designed to deliver enhanced mobile broadband with higher data rates up to $20$~Gbps. 
The growth in data rates is made possible partly by carrier aggregation, which has increased the total supported bandwidth from $100$~MHz used in long-term evolution advanced (LTE-A) to over $1$~GHz in 5G-NR\cite{slim_2019}. 
The increase in bandwidth leads to a significant challenge in linearization of power amplifier (PA) over a wide bandwidth. 
Digital pre-distortion (DPD) has been widely adopted to improve PA linearity\cite{dpd_thesis}. 
The conventional DPD is mainly based on the Volterra series\cite{volterra_2012} or its simplified versions like memory polynomial\cite{mem_poly_2006}. 
Recently, researchers have introduced a deep neural network (DNN) framework for DPD operation\cite{arvtdnn, cnn_dpd, etdnn}. 
Most DNN-based DPD systems are implemented using software models written using popular machine learning libraries in python. 
Although these models perform well on offline data, adopting them to real-time hardware designs such as FPGA is challenging due to timing and resource constraints. 
These challenges could be addressed by using either a radio-frequency (RF) analog accelerator\cite{udara_cnn} or a digital accelerator\cite{tpu_2017}. 
This paper presents a novel modular implementation of a one-dimensional convolutional neural network (1D-CNN) that can operate in real-time. 
The modular nature of the architecture increases the adaptability of the system to different neural network configurations. 
The synthesis results show that the presented DNN-DPD system can operate in greater than $100$ MHz bandwidth which can be used to pre-distort wideband orthogonal frequency-division multiplexing (OFDM) signals used in 5G-NR.

\section{DNN-based DPD}


In practice, applying DNN to linearize RF power amplifier in radio transmitter demands special attention, due to the strict latency and resource constraints. Thus several modifications to conventional DNN should be made to adopt them to DPD. 
A commonly used simple approach for this is the real-valued time-delayed neural network (RVTDNN)\cite{dpd_book}. 
It was further modified by the augmented real-valued time-delayed neural network (ARVTDNN)\cite{arvtdnn} which extends the input layer by adding pre-calculated envelope and its higher-order terms. 

\begin{figure}
    \centering
    \includegraphics[width=0.5\textwidth]{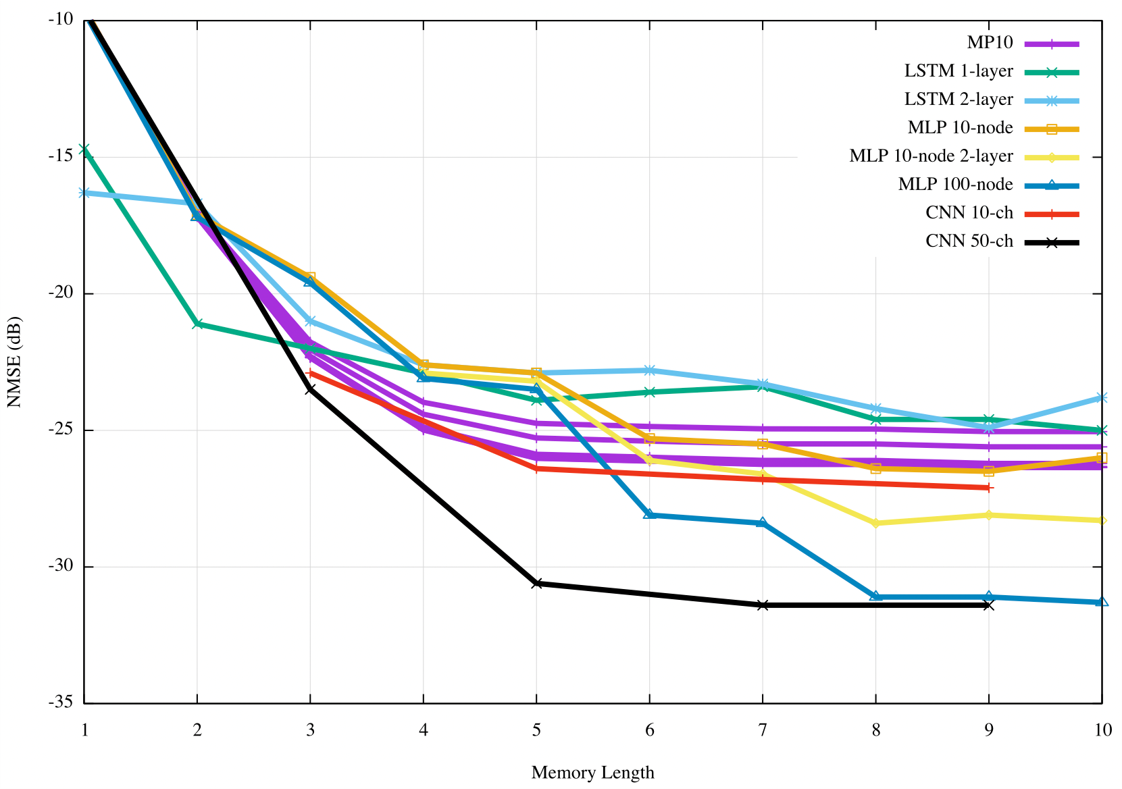}
    \caption{NMSE performance of different NN architectures for PA modelling.}
    \label{fig:toshi_nn_results}
\end{figure}

Convolution neural networks (CNNs) have been also investigated for DPD application\cite{cnn_dpd}, where the input data and its envelop terms are stacked together for two-dimensional CNN. 
Our simulation results in Fig.~\ref{fig:toshi_nn_results} shows that one-dimensional (1D) CNN can outperform several other configurations, including RVTDNN based on multi-layer perceptron (MLP), recurrent neural networks based on long short-term memory (LSTM), and conventional memory polynomial.
Motivated by this result, we investigate a hardware-friendly 1D-CNN design for the DPD application. 

\section{Design of modular 1D-CNN architecture}

\begin{figure}
    \centering
    \includegraphics[width=0.45\textwidth]{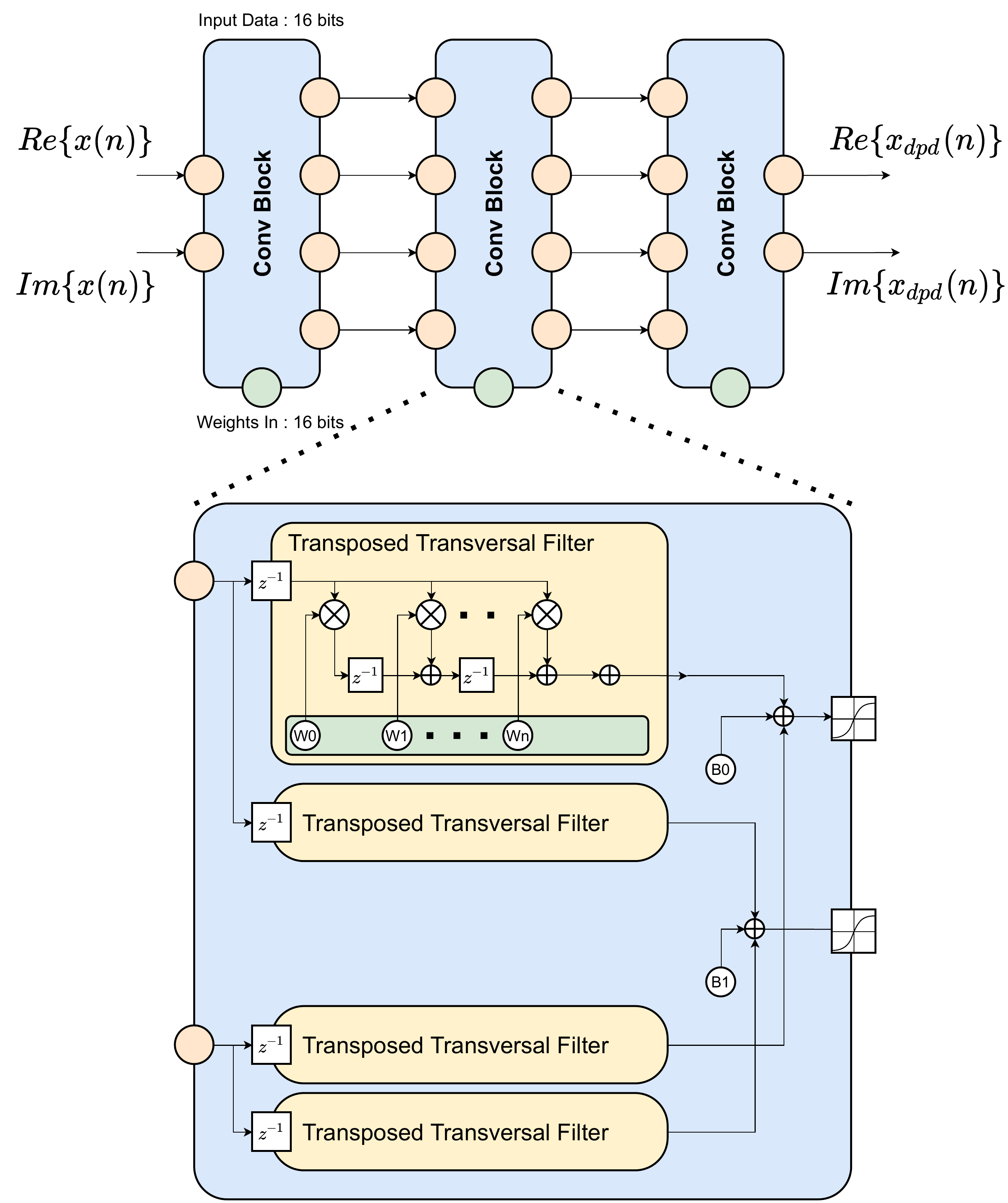}
    \caption{System architecture of 1D-CNN processor.}
    \label{fig:overall}
\end{figure}

Our architecture adopts the modularity and parametric design principles for the adoptability of our design to other DNN architectures. 
The overall 1D-CNN architecture is shown in Fig.~\ref{fig:overall}. The ConvBlock is a parameterized implementation of the 1D convolution defined 
with a kernel size $K$, an input channel $C_\mathrm{in}$ and an output channel $C_\mathrm{out}$ for specification of kernel weights and biases.
A system that converts an input with $C_\mathrm{in}$ channels to an output with $C_\mathrm{out}$ channels should have $C_\mathrm{in} \times C_\mathrm{out}$ number of 1D filters. 
The design uses the `generate' construct in Verilog to allow the design to support a user-specified number of input/output channels.

The 1D filters are implemented using the transposed finite-impulse response (FIR) filter structure\cite{oppenheim_dsp} since it is known that the transposed form filters can outperform the direct form filters when the filters are large\cite{xilinx_transposed}.
In the transposed FIR filters, multipliers can be avoided entirely if the coefficients are powers of two (PoT)\cite{xilinx_transposed} which leads to more hardware-friendly designs. 

In order to update weights without modifying any design files, the weights are declared in a separate file and included in the Conv1d implementation. 
This design pattern requires resynthesizing every time the weights are updated. 
Alternatively, weights can be stored in registers which can be updated by a processor.  

The design supports various nonlinear activation function such as sigmoid. 
Our experiment showed that rectified linear unit 6 (ReLU6) outperforms sigmoid activation in 1D CNN-based DPD systems. 
The ReLU6 implementation is also hardware-friendly compared to most other activation functions as it just needs two cooperators.


\section{Verification using co-simulation}

\begin{figure}[t]
    \centering
    \includegraphics[width=0.35\textwidth]{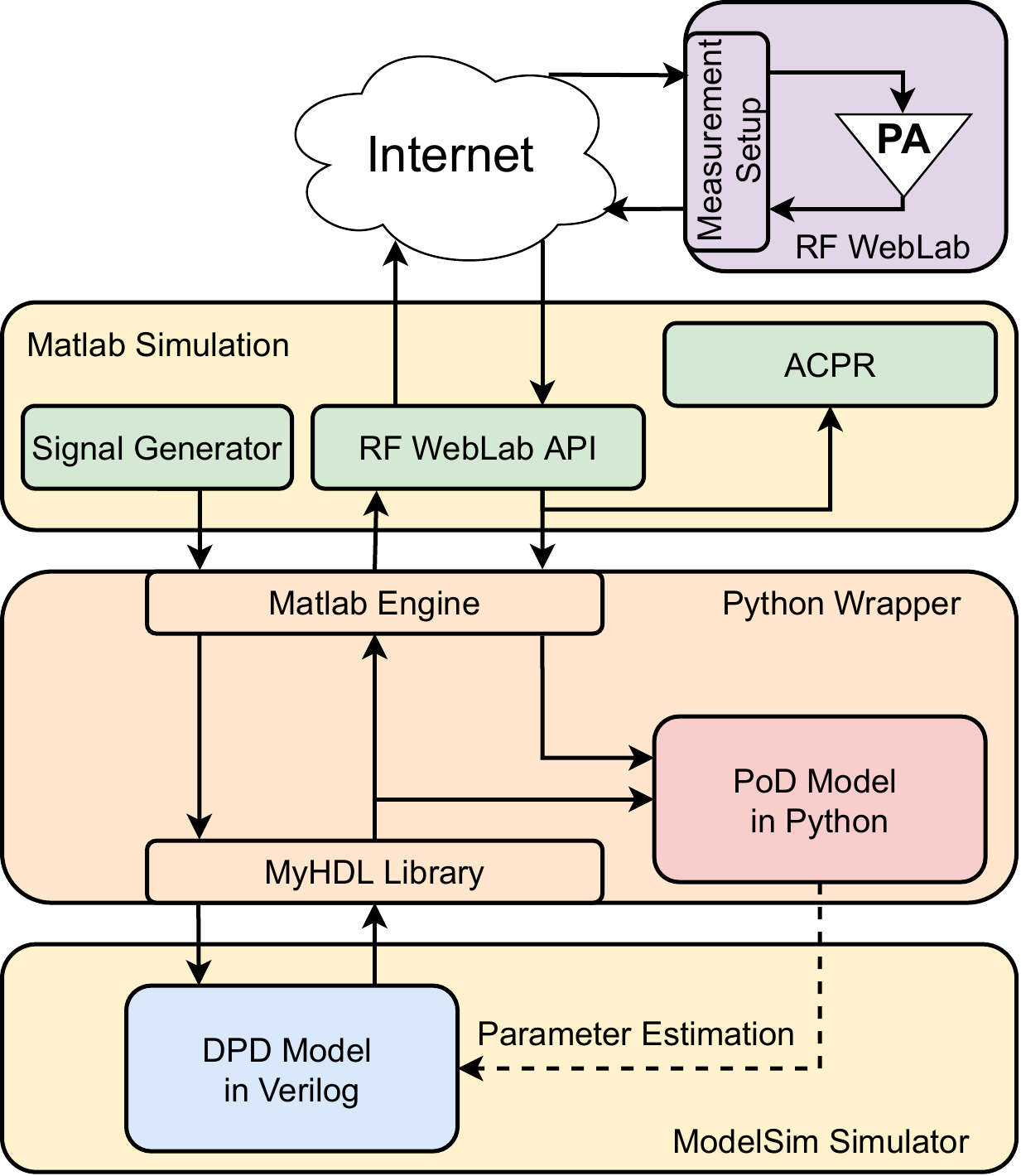}
    \caption{Verification architecture.}
    \label{fig:verification}
\end{figure}

A co-simulation strategy is used to verify the design. 
The verification setup uses the ModelSim as the register-leverl transfer (RTL) simulator and Python to communicate with RFWebLab\cite{rfweblab} and simulate the software-based 1D CNN. 
The complete verification architecture is shown in Fig.~\ref{fig:verification}. 
MyHDL\cite{myhdl} and Matlab Engine are used through Python libraries that communicate with ModelSim and Matlab. 
OFDM signals are generated in MATLAB and provided as the input to the DPD system running in ModelSim simulator. 
The output is sent to RFWebLab, an online accessible testbed with a real PA. 
The output from PA is returned to MATLAB, to measure the improvement. 

\begin{figure*}[t]
    \centering
    \includegraphics[width=\textwidth]{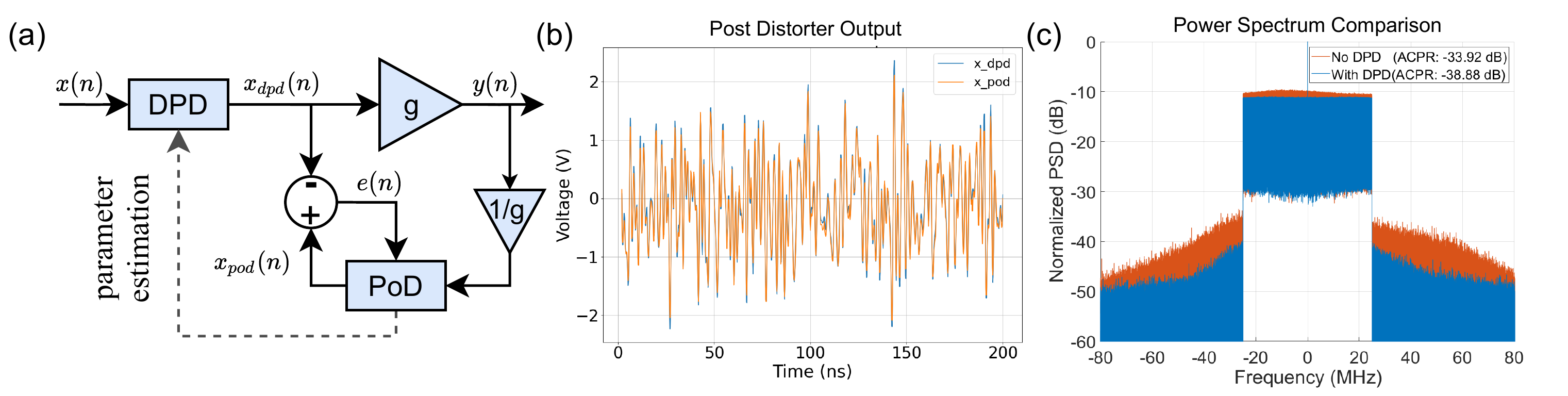}
    \caption{DPD verification using co-simulation: 
    (a) In-direct learning architecture; 
    (b) Post distorter output ($x_{pod}(n)$ comparison with PA input ($x_{dpd}(n)$) after training;
    (c) Power spectral density of the transmit signal without using DPD (shown in red) which has an ACPR of $-33.92$~dB and transmit signal after using DPD (shown in blue) which has an ACPR of $-38.88$~dB.}
    \label{fig:dpd_verify}
\end{figure*}

The DPD is trained using the indirect learning architecture (ILA) shown in  Fig.~\ref{fig:dpd_verify}(a). 
A post-distortion (PoD) model is used in ILA, which has the same architecture as the DPD model. 
The PoD model is trained using the PA output $y$ and PA input $x_\mathrm{dpd}$. 
The error is calculated by $e(n) = x_\mathrm{pod}(n) – x_\mathrm{dpd}(n)$. 
The PoD model is trained until the normalized mean squared error is minimized. 
Fig.~\ref{fig:dpd_verify}(b) shows that PoD model output $x_\mathrm{pod}(n)$ closely predicts the PA input $x_\mathrm{dpd}(n)$ when the error is minimized. 
After the training, the parameters of the PoD model are passed on to the DPD model.



We use $50$~MHz OFDM signals to test the DPD performance in terms of adjacent channel power ratio (ACPR)\cite{acpr_2012}. 
The experimental results showed that DPD improves the ACPR from $-32.92$~dB to $-38.88$~dB. 
The power spectral density (PSD) comparison with and without the DPD is also shown in Fig.~\ref{fig:dpd_verify}(c).


The hardware design is synthesized targeting the Xilinx ZCU104 FPGA board. 
Table~\ref{tbl:synth} shows the timing and area reports of the design synthesis for different configurations. 
All configurations use the fixed point number representation with $16$ total bits and $10$ fractional bits. 
The ACPR results shown in Fig.~\ref{fig:dpd_verify}(c) is taken from the design with two hidden layers, $20$ neurons per layer, and a kernel size of $5$. 
The performance could be increased by increasing them at the cost of more hardware resources. 
The timing results in Table~\ref{tbl:synth} show that our design meets the requirement to run at least $100$~MHz for $50$~MHz bandwidth. 

\begin{table}[th]
\begin{threeparttable}
\caption{Timing and resource utilization of hardware design for different network configurations}
\label{tbl:synth}
\begin{tabular}{l l l| l |l l l}
\hline
\multicolumn{3}{c|}{Design} &
 {Timing} &
  \multicolumn{3}{c}{Area} \\ \hline
\begin{tabular}[c]{@{}l@{}}hidden \\ layers\end{tabular} &
  \begin{tabular}[c]{@{}l@{}}neurons\\ per layer\end{tabular} &
  \begin{tabular}[c]{@{}l@{}}kernel \\ size\end{tabular} &
  \begin{tabular}[c]{@{}l@{}}Critical \\ path (ns)\end{tabular} &
  LUTs &
  Regs &
  DSPs \\ \hline
1 & 20 & 5 & 1.541 & $\sim$4k  & $\sim$1k  & 400 \\ 
1 & 20 & 9 & 1.541 & $\sim$4k  & $\sim$1k  & 400 \\ 
2 & 20 & 5 & 3.218 & $\sim$55k & $\sim$46k & 914 \\ 
1 & 40 & 5 & 1.541 & $\sim$5k  & $\sim$7k  & 510 \\ 
2 & 40 & 5 & 4.040 & $\sim$76k & $\sim$59k & 1k  \\ \hline
\end{tabular}
\begin{tablenotes}
\footnotesize \item The synthesis is targeted for Xilinx MPSoC FPGA in ZCU104 board which has 230k LUTs, 460k Regs and 1.7k DSP blocks
\end{tablenotes}
\end{threeparttable}
\end{table}



\section{Conclusion}

This paper presents a hardware-friendly modular architecture of a 1D-CNN-based DPD system. 
Due to the modular nature and parametric design, this architecture is easily adaptable to different DNN configurations to accommodate different timing and resource constraints. 
The design is verified using a co-simulation strategy that measures the ACPR improvement for an FPGA. 
The performance of the DPD system can be approximated in significantly less time since it avoids the bit-stream generation time. 
The initial experimental results showed around a $5$~dB reduction in ACPR for two-layer 1D-CNN with $20$ neurons per hidden layer. 
This performance can be further improved by using a deeper/wider network configuration at the cost of increased resource utilization. 
A software-based post-distortion model is used to extract the parameters for the DPD model. 
In future work, we will investigate direct learning architecture to update the DPD model parameters on the fly. 
The current design uses DSP blocks in the FPGA to perform multiplication. 
In the future, we will investigate PoT approximations for the model parameters and the improvement in resource utilization by replacing DSP blocks with shift and addition operations.






%
\bibliographystyle{IEEEtran}
\bibliography{dpd_ref.bib}

\end{document}